# A Deep Learning Perspective on Connected Automated Vehicle (CAV) Cybersecurity and Threat Intelligence

*Manoj Basnet[1], Mohd. Hasan Ali[2]*

[1,2] Department of Electrical and Computer Engineering, The University of Memphis, Memphis, TN  [1]mbasnet1@memphis.edu, [2]mhali@memphis.edu

**Abstract**

The automation and connectivity of CAV inherit most of the cyber-physical vulnerabilities of incumbent technologies such as evolving network architectures, wireless communications, and AI-based automation. This book chapter entails the cyber-physical vulnerabilities and risks that originated in IT, OT, and the physical domains of the CAV ecosystem, eclectic threat landscapes, and threat intelligence. To deal with the security threats in high-speed, high dimensional, multimodal data and assets from eccentric stakeholders of the CAV ecosystem, this chapter presents and analyzes some of the state of art deep learning-based threat intelligence for attack detection. The frontiers in deep learning—namely Meta-Learning and Federated Learning-- along with their challenges have been included in the chapter. We have proposed, trained, and tested the deep CNN-LSTM architecture for CAV threat intelligence; assessed and compared the performance of the proposed model against other deep learning algorithms such as DNN, CNN, LSTM. Our results indicate the superiority of the proposed model although DNN and 1d-CNN also achieved more than 99% of accuracy, precision, recall, f1-score, and AUC on the CAV-KDD dataset. The good performance of deep CNN-LSTM comes with the increased model complexity and cumbersome hyperparameters tuning. Still, there are open challenges on deep learning adoption in the CAV cybersecurity paradigm due to lack of properly developed protocols and policies, poorly defined privileges between stakeholders, costlier training, adversarial threats to the model, and poor generalizability of the model under out of data distributions.

## Contents

# 1. INTRODUCTION

The connected and autonomous vehicle (CAV) is the next-generation mobility service—powered by intelligent automation and robust communication--aimed at replacing human-maneuvered vehicles with the software agent matching or even exceeding the human-level intelligence, control, and agility with the least decision errors possible. US National Safety Council (NSC) [1] reported a 24 % spike in roadway death rates from 2019 despite miles driven dropping 13%, the highest increment in 96 years in the US since 1924. Most of the time it's from human error. NSC estimated 4.8 million additional roadway users were seriously injured with an estimated cost of $474 billion in the year 2020. Figure 1. represents the preliminary motor vehicle death estimates from 2018 to 2021 [2]. The next generation of transportation and mobility envisions the safe, reliable, agile, automated, trustworthy, and service-based mobility architecture. The architecture should be able to eliminate human errors by using intelligent decision-making software agents based on the situational and

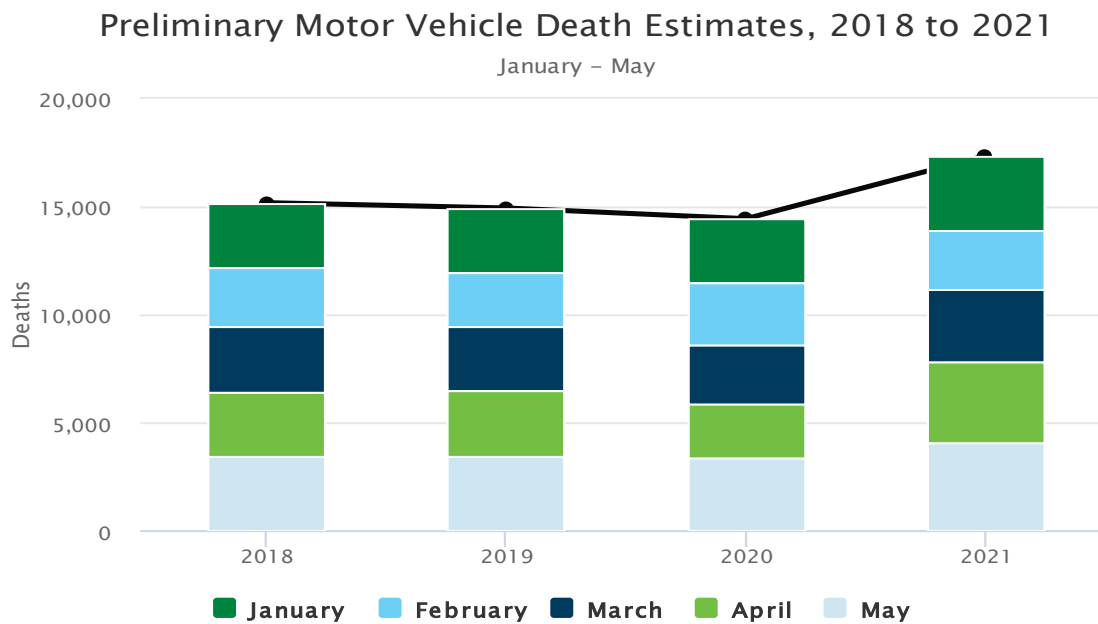

**FIGURE 1. Preliminary death estimates from 2018 to 2021 [2].**

behavioral information collected by sensors and transceivers through communication. Apart from that, the service-based architecture removes the concept of vehicle ownership and includes more diversity such as disabled, elderly people. CAV is the evolving technology to achieve the envisioned goals of future mobility and transportation.

Commutation vehicles nowadays are not merely electromechanical entities, but also complex software agents with electronics [3]. Connected means the vehicle exchange the data between the systems and networks (to other vehicles and infrastructures) for predictive maintenance, dynamic insurance policy, passenger information, fleet management, comfort, situational and behavioral awareness [4]. Fully autonomous means the vehicle conducts dynamic driving tasks automatically in real-time without the driver's intervention [5].

The connected vehicle generates 25 GB of data per hour even at the lower level of autonomy. The integration of RADAR, LIDAR, Camera, Ultrasonics, Motion sensors, GNSS, IMU to the vehicle can generate 40 Gbit/s data leading up to 380 TB to 5100 TB of data just in a year [6]. This wealth of high volume, high-speed data needs gigantic storage, intense computation, and astute processing. As the data volume vehemently upsurges, the privacy and security of software, hardware, and data itself become very critical.

Increased connectivity elevates the attack surfaces of the CAV, while automation lacks the sophisticated human-level agility and intelligence for threat mitigation. The inherent vulnerabilities come from the untested supply chain such as hardware, software, and infrastructures.

Deep learning has been unprecedently successful in deciphering the complex nonlinear spatiotemporal pattern of highly stochastic data. Given the volume, veracity, and velocity of the data, deep learning could be handy in designing cyberattack detection and mitigation in the CAV environment.

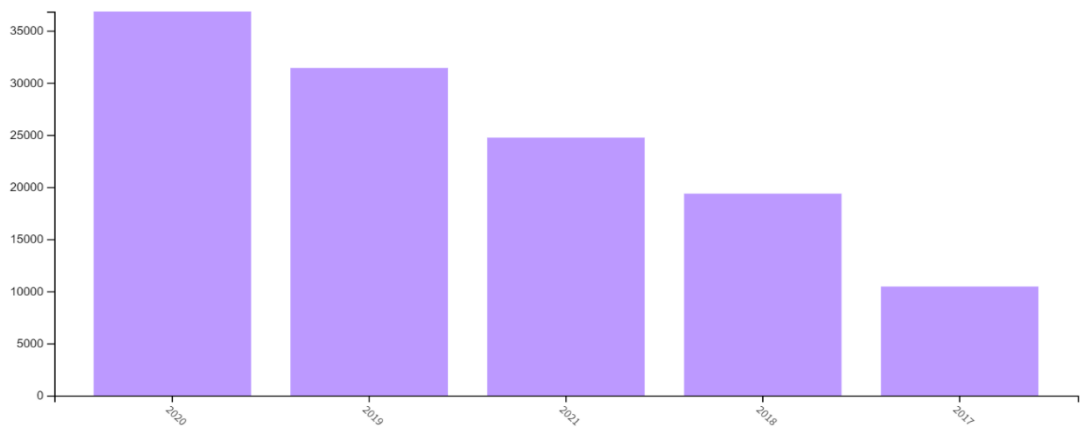

**FIGURE 2. The number of publications on connected automated vehicles over the last five years.**

Figure 2. shows the current trends of publications queried under "(((ALL=(Connected Vehicle )) OR ALL=(Automated vehicle)) AND ALL=(Cybersecurity)) OR ALL=(Deep learning)" in the web of science for 2017 through 2020. It resulted in 127,042 publications so far with growing interest per year.

This chapter deals with the deep learning perspective on CAV cybersecurity and threat intelligence. Also, We proposed novel end-to-end deep CNN-LSTM based computational intelligence for cyberattack detection and classification in the CAV ecosystem. The proposed model has been successfully trained, tested, and evaluated on the CAV-KDD dataset and compared against other deep learning models such as Deep Neural Network (DNN), Convolutional Neural Network (1D-CNN), Long-Short Term (LSTM). The proposed model outperformed the aforementioned deep learning models in terms of various performance metrics along with the increased model complexity.

## 2. CAV TECHNOLOGICAL ENABLERS: AUTOMATION AND CONNECTIVITY

The key technological enablers for CAV are Automation and Connectivity. Vehicle driving automation system performs part or all of the dynamic driving tasks (DDT). The Society of Automotive Engineers (SAE) defined the six levels of automation for vehicles ranging from no automation (Level 0) to full automation (Level 5) on its 2021 release [7]. Table I summarizes the level of driving automation system (ADS), the role of the user, and the role of driving automation systems. In Levels 0-2, the driver drives the entire or part of DDT, while level 3-5 ADS performs the entire DDT upon engagement.

**TABLE I.** EVOLUTION OF DRIVING AUTOMATION [7, P. 30]

| Level of Driving Automation | Role of User | Role of Automation System |
|---|---|---|
| Level 0- No driving Automation | Driver always performs the entire DDT. | No DDT at all, safety warning and some other momentary emergency intervention in some cases |
| Level 1- Driver Assistance | Driver supervises the driving automation, performs all other DDT, determines when to engage or disengage the driving automation, and controls the entire DDT upon requirement or desire. | ADS, while engaged, executes either longitudinal or lateral vehicle motion control subtask and hand over the control to the driver upon immediate intervention. |
| Level 2-Partial Driving Automation | Driver supervises the driving automation, performs all other DDT, determines when to engage or disengage the driving automation, and controls the entire DDT upon requirement or desire. | ADS, upon engagement, execute both longitudinal and lateral vehicle motion control subtask and disengages upon driver request. |
| Level 3-Conditional Driving Automation | Driver, while ADS is not engaged, verifies the operational readiness of the ADS, and decides whether to engage ADS or not. Once ADS got engaged, the driver becomes DDT fallback-ready user. Though ADS is engaged, a user is always receptive to user intervention requests and DDT performance-relevant failures. Moreover, a user is always there for risk assessment and ADS disengagement. | ADS, while disengaged, permits the operations only within its operational design domain (ODD). Users can immediately take control upon the request. ADS, while engaged, perform the entire DDT within its ODD, transmits intervention requests to the DDT fallback-ready user when ODD limits are reached or when DDT performance-relevant failures are detected. |
| Level 4-High Driving Automation | Driver, while ADS is not engaged, verifies the operational readiness of the ADS, and decides whether to engage ADS or not. Once ADS got engaged, the driver becomes a passenger. Upon ADS engagement, the passenger does not perform DDT or DDT fallback or risk assessment. A passenger may disengage the ADS and become the driver and perform the DDT after ODD limits have been reached. | ADS, while disengaged, permits the operations only within its operational design domain (ODD). ADS, while engaged, perform the entire DDT within its ODD, It may prompt the passenger to resume operations of the vehicle near the ODD limits. Also, It performs DDT fallback transitions upon user request to achieve minimal risk or under DDT-relevant system failures or reaching the ODD limits. It may cause some delay for the disengagement. |
| Level 5- Full Driving Automation | Driver, while ADS is not engaged, verifies the operational readiness of the ADS, and decides whether to engage ADS or not. Once ADS got engaged, the driver becomes a passenger. Upon ADS engagement, the passenger need not perform DDT or DDT fallback or risk assessment. A passenger may disengage the ADS, becomes the driver, and performs the risk assessment. | ADS, while not engaged, permits engagement of ADS under driver manageable all road conditions. ADS, while engaged, perform all the DDT, performs automatic DDT fallback transitions to minimal risk condition upon DDT performance -relevant failures or user request to achieve minimal risk condition. ADS got disengaged only when it achieves minimal risk conditions, or a driver is performing the DDT. It may cause some delay for the disengagement. |

Figure 3. shows the overall connectivity architecture with a wireless network interface, physical interface, and In-vehicle network in between. CAV is co-evolving with next-generation network architectures and communication protocols. It can exploit the latency, speed, and bandwidth of recent cellular communication such as 5G. The recent advancement in Wi-Fi 6 could be used in the place of high user density. Also, inbuilt GPS has been widely used for navigation. Moreover, Bluetooth, RFID, ZigBee, and V2X communication have extended the range of connectivities and applications. The in-vehicle network has mainly a high-speed infotainment system for information dissemination and entertainment; and a Powertrain network for core functionalities of the vehicle. These are mostly composed of electronic control units (ECUs) connected through local control area network (CAN) buses. Physical network interface provides ports to connect the phone, USB, CD, AUX from the infotainment system, and ODB-II from the powertrain system. This ODB-II can be extended via physical or wireless network interfaces to OEM, drivers, and computerized intelligence in the vehicle.

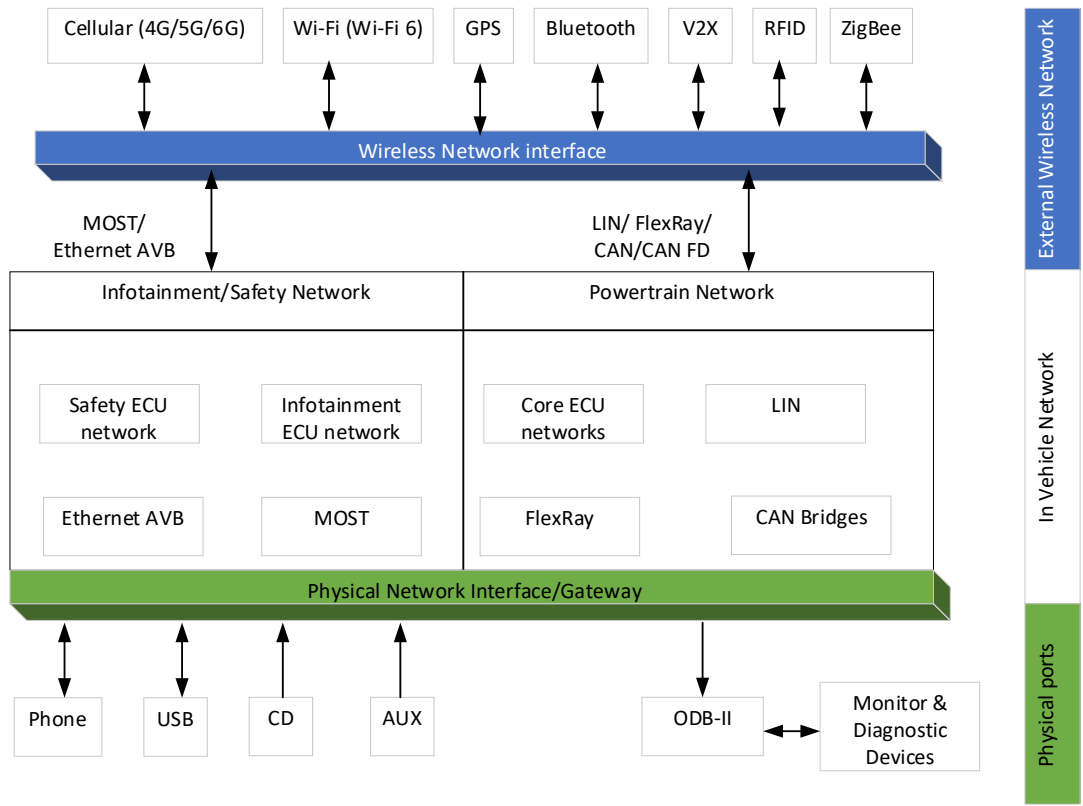

**FIGURE 3. Connectivity of the CAV adapted from [8].**

The hitherto advancement in connectivity and automation has enabled the following vehicle functionalities as per the US department of transportation (DoT) [9].

a) Collision warning: forward collision warning, lane departure warning, rear cross-traffic warning, and blind-spot warning

b) Collision intervention: automatic emergency braking, pedestrian automatic emergency braking, rear automatic braking, and blind-spot intervention.

c) Driving control assistance: adaptive cruise control, lane-centering assistance, and lane-keeping assistance.

d) Additional systems: Automatic high beams, backup camera, and automatic crash notification.

## 3. CAV THREAT LANDSCAPE AND THREAT INTELLIGENCE

Threat intelligence begins with identifying the assets and then finding the weighted utility to the assets i.e., threat landscape. Assets are the entities that have specific utilities and hence add values to the system. The value comes from the cost of creating it and competition to make it easily available. Therefore, from a game-theoretic perspective, there is always competition to exploit the utility i.e. risk for biased usage which in turn creates vulnerabilities. An attacker can exploit the vulnerabilities by using social engineering and reverse engineering. The electromechanical vehicle while adopting the evolving network architecture and automation—so-called CAV-- migrates all the vulnerabilities related to the processes, protocols, supply chains, software from the incumbent technologies. Furthermore, CAV has vulnerabilities or risks originated from communication, automation, IT, OT, and physical system. Here, we will briefly explain the cyber vulnerabilities of low-level sensors [10] and vehicle control modules.

## 3.1. IN-VEHICLE (LOW-LEVEL SENSOR) CYBER VULNERABILITIES

1. GPS: The transparent architecture of GPS, its open standard, and free accessibility are the main reasons for generating spoofing and jamming attacks on GPS.
2. Inertial measurement units (IMUs): These units provide velocity, acceleration, and orientation data by using accelerometers and gyroscopes. The gyroscope and inclination sensors measure the road gradient and adjust the speed accordingly for safe maneuvering. The spoofed data can generate the false control signal for speed control. Also, the jamming of the sensors may disrupt the vehicle's autonomous speed adjustment.
3. Engine control sensor: These sensors monitor the dynamics of the engine such as temperature, airflow, exhaust gas, engine knock and are connected to CAN.
4. Tire Pressure Management System (TPMS): TPMS has not been used in decision-making but is physically accessible to outsiders.
5. LiDAR sensors: are used to generate the 3D map of the vehicle's environment for localization, obstacle avoidance, and navigation. LiDAR sensors can be fooled by Laser beams.
6. Cameras (stereo- or mono-vision) and infrared systems: These are used for static and dynamic obstacle detection, object recognition, and 360-degree information with other sensor fusion. Cameras contain the charge-coupled device (CCD) or complementary Metal Oxide Semiconductor (CMOS) sensor that can be partially disabled from a 3- meter distance using low-powered lasers.

## 3.2. VEHICLE CONTROL MODULES

All modern vehicles use engine control units (ECU) for the acquisition, processing, and control of electronic signals. ECUs are roughly categorized into powertrain, safety systems, body control, and data communications. The powertrain is the brain of ECU that controls transmissions, emissions, charging systems, and control modules. Safety systems are responsible for collision avoidance, airbag deployment, active braking, and so on. Body control controls the electric windows, AC, mirrors, immobilizer, and locking. Data communications control the communication between different communication modules. The networking of ECUs can be done through either CAN buses or FlexRay. The key ECUs in CAV in descending orders of importance are as follows [10], [11].

a) Navigation control module (NCM)
b) Engine control module (ECM)
c) Electronic brake control module (EBCM)
d) Transmission control module (TCM)
e) Telematics module with remote commanding
f) Body control module (BCM)
g) Inflatable restraint module (IRM)
h) Vehicle vision system (VVS)
i) Remote door lock receiver
j) Heating, ventilation, and air conditioning (HVAC)
k) Instrument panel module
l) Radio and entertainment center

## 3.3. SECURITY ANALYSIS OF CAV THREATS

CAV can have around 100 million lines of code across 50-70 ECUs. As the number of lines of code grows It's infeasible to perform careful security implications. Some security incidences and their analysis are presented here [4].

a) Remotely control a vehicle: The attacker exploits the vulnerability in the cellular system and lands on the infotainment system. In most of the vehicle, the infotainment system has a driver with information such as service schedules, tire pressure and so on. The infotainment system has a connection with the CAN bus that connects all the ECUs. Therefore, it is possible to enter through the infotainment system and inject or spoof malicious signals. For eg: ECUs controlling steering or brake.
b) Disable the vehicle: exploiting flaws in authentication, authorization, and access control in smart devices and apps to activate AC, windows, windshield to drain the battery.
c) Remotely unlock the vehicle/theft: exploit known vulnerabilities in keyless entry system using SDR. Hackers remotely unlocked the car door and started the engine in the Mercedes Benz-E class in 2020. The manufacturer generally uses symmetric keys between the key fob, entry system, and ignition keys. An attacker can sniff the radio frequency between key fob and entry system either by brute force or as a man-in-the-middle attack. Later the symmetric key can be compromised by replay attack or reverse engineering. The problem became humongous when some leading vehicle manufacturers use the same master cryptographic keys along the model line.
d) Safety conditions: Panic attacks such as Mobileye and Tesla X hack fooled the autopilot system to trigger the brakes and steer into an oncoming vehicle.
e) Vehicle tracking/monitoring: extract patterns or fingerprints from the data.
f) Weaponizing the vehicle
g) Malware: bots for crypto-jacking or DDoS.
h) Ransomware could be a huge problem to be dealt with in the future CAV.
i) Distribution of illicit goods.

## 3.4. ATTACK SURFACES

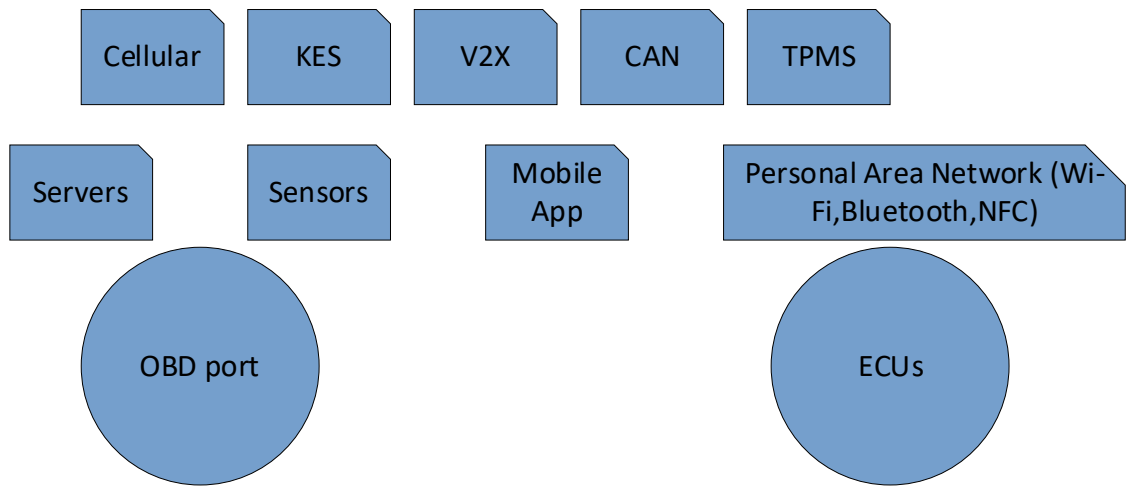

**FIGURE 4. Key attack surfaces of CAV.**

The orthodox electro-mechanical vehicles while adapting the evolving network architecture, communications and AI-powered automation migrate all the vulnerabilities of incumbent technologies. With the elevated sophistication, CAV also inherits elevated attack surfaces and attack vectors. These attack vectors are the specific methods, paths, or processes through which the CAV can be exploited. Insider threats such as Levandowski trade secret trial between Waymo and Uber [12], Cyberattack into V2X communications [13], Sensor spoofing and exploitation [14], Dumpster diving for data: acting as a honeypot, Supply chain, and third-party risks are some of the prominent threats in CAV ecosystem. Figure 4. enlists the prime attack surfaces.

## 3.5 ORGANIZATIONAL RISKS TO CAV ECOSYSTEM

The organizational risks imposed on the CAV ecosystem are well documented in [15]. The convergence of IT security, OT security, and physical security is a challenging issue in any cyber-physical system including the CAV ecosystem. The interconnections, interactions, and co-impacts of attack on these eccentric systems should be analyzed and evaluated. Dealing with big data (high volume, high speed, high variety) and extracting inferences in the CAV ecosystem needs high computational capacity, storage, and processing to deal with the multimodal data from different sensors. The data communication between multiple nodes, servers, and systems impose data security and privacy risk. The divergent nature of stakeholders such as different vendors of CAV, OEM, ITS, V2X, and its data privacy policy might not allow CAV actors to collaborate in threat detections and mitigations. The cyber-physical security protocols, enterprises policies, and regulations still have to go long way in the CAV ecosystem.

# 4. CAV THREAT MITIGATION: ANOMALY DETECTION AND CLASSIFICATION WITH DEEP LEARNING

"Deep Learning is building a system by assembling parameterized modules into a (possibly dynamic) computation graph and training it to perform a task by optimizing the parameters using a gradient-based method" as quoted by Yann Le Cunn, ACM Turing awardee and a pioneer in deep learning in AAAI-20 event [16]. Graphs can be defined dynamically by input-dependent programs. Output computation may not be necessarily the feedforward, it might be some minimizing energy functions (inference model) [17]. The designer has complete freedom to choose learning paradigms such as supervised, reinforced, self-supervised/unsupervised, and objective functions such as classification, prediction, reconstruction. Often limitations of supervised learning are mistakenly seen as limitations of deep learning. If a cake is intelligence, self-supervised learning is the bulk of the cake, supervised learning is the icing of the cake and RL is the cherry on the cake. The next revolution in AI won't be supervised nor reinforced [18].

Deep learning has been an exciting paradigm for anomaly detection and classification in various cyber-physical realms such as industrial control systems, smart grids, SCADA-controlled systems, and so on [19]. Now specific state of art applications of deep learning in CAV cybersecurity is summarized. In [20], Generative Adversarial Network-based IDS has been used to detect the anomaly in ECU by analyzing the CAN message frame specifically the message identifier and its frequency. The dataset was recorded from the OBD-II port of an undefined vehicle. The authors modified the firefly algorithm to find the optimal structure of the Generator network. Finally, they claimed the superior accuracy of the proposed model to the PSO- and GA- optimized GAN. However, the paper does not have much information regarding the training time, data size, data samples, computational complexity, and so on.

In [8], a Deep learning-based LSTM autoencoder has been implemented to design IDS for CAN and central network gateways using car hacking and UNSW-NB15 datasets respectively. The statistical features such as total count, mean, standard deviation are extracted from the network packets. The proposed model claimed to outperform some of the decision tree and SVM-based classifiers. It's unlikely to claim DL model can detect zero-day attacks since the supervised ML model cannot detect and classify the data that have never been trained.

In [21], authors use GAN for designing IDS capable of learning unknown attacks in the in-vehicle network. They extracted the CAN bus data for normal and attack by using raspberry pi and simple hardware in the OBD-II port. Instead of converting all the CAN data to an image (make real-time detection at stake due to increased processing), only CAN IDs are converted into the image by using one-hot encodings. For training, the first discriminator uses the normal and abnormal CAN images extracted from the actual vehicle, while the second discriminator uses normal and random noise. The generator and discriminator compete to increase their performance and the second discriminator can detect fake images similar to real CAN images. The proposed model, however, has used only CAN IDs as the main feature to identify the attack from non-attack. Converting data into images hinders the real-time detection of IDS. Also, the model can't detect operational flaws from the attack.

In [22], ResNet-inspired DCNN has been used for sequence learning of broadcasted CAN IDs using the same dataset as in [21]. However, they are more interested in finding the pattern in the sequence of IDs rather than individual IDs. 29 bits IDs are recorded for every 29 consecutive IDs forming a 29x29 grid image ready to go into the DCNN and correspondingly labeled as an attack or no attack. They claimed that the DCNN seems to be more efficient in sequence learning than LSTM for this problem. This model needs high computational power and cannot detect unknown attacks.

Yu [23] proposed a novel self-supervised Bayesian Recurrent Autoencoder to detect adversarial trajectory in Sybil attacks targeting crowdsourced navigation systems. It uses time-series features of vehicle trajectories and embeds the trajectories in a latent distribution space as multivariate random variables using an encoder-reconstructor. This distribution is used to reconstruct the authentic trajectories and compared with the input to evaluate the credibility score. The author claimed that this model improves the baseline model by at least 76.6 %.

## 5. FRONTIERS IN DEEP LEARNING (ADVANCEMENT AND FUTURE)

The challenges of deep learning: Supervised models need extensive labeling of data while reinforcement learning needs a very large number of interactions. Very slight modifications in fewer pixels and even a small change in rules in the environment can err the model. The inefficacy of the deep learning-based models is rooted in the assumption of "independent and identically distributed (i.i.d)" data. This assumes that the training data capture all the stochasticity of the real dynamic environment and observations are independent of each other. To capture the dynamics of changing environment, the learner model should evolve accordingly. Deep learning models are data-hungry; future deep learning should be envisioned to learn with fewer samples and fewer trials. For that model should correctly and broadly understand the environment before learning the tasks. Deep learning models are very poor at abstraction and reasoning needs humongous data just to learn a simple task. Symbolic AI has proven to be much better at reasoning and abstraction. Deep learning models are good at providing end-to-end solutions but miserable at breaking them down into interpretable and modifiable subtasks.

Currently, deep learning is said to have achieved system I natural intelligence i.e., just achieving associative or mapping intelligence. For example, a human driver navigates to the neighborhood with visual cues that have been used a hundred times before without looking up at the direction or map. Also, while navigating to the new environment, he could use a map, direction, reasoning, logic to get to the destination. The first one is the system I cognition while the second is the system II cognition [24].

The pioneer deep learning scientists pointed out the following roadmaps for the future AI to be more conscious (system II cognition) at NeurIPS 2019:

- Handling the out of distribution (o.o.d) nonstationarity in the environment
- Systematic generalization
- Consciousness prior
- Meta-learning and localized change hypothesis for causal discovery
- Cosmopolitan DL architectures

## 5.1. META-LEARNING

When we start learning some new tasks, we merely start from scratch rather we try to use prior experiences ($\theta_i \in \Theta$) from the prior known tasks ($t_j \in T$). Where $\Theta$ is the discrete, continuous, or mixed configuration space and T is the set of all known tasks. For example- how a human driver can easily drive in a completely new environment. Along the way of learning specific tasks human brain also learns how it is learning. These prior learning experiences from the tasks, if applied to learn new tasks, could bring one step closer to get the cognitive power of system II. As a result, this new model would learn the new tasks with sparse data in a short time. Meta-learning or learning to learn is the science of transferring learning experiences, metadata, from the broader tasks to learn a new task with the least information in the least possible time [25]. The meta-data embody the prior learning tasks and the learned models in terms of exact algorithm configurations, hyperparameters, network architectures, the resulting performance metrics such as accuracy, training time, FAR, F1-score, prior weights, and measurable properties of tasks (meta-features). Once meta-data is collected, a machine needs to extract and transfer knowledge of the meta-data to search for the optimal models to solve the new tasks. Paper [25] explains how meta-learners learn from the model evaluations such as task-independent recommendations, configurations of space design, and its transfer techniques including surrogate models and warm started to multitask learning. As opposed to the base learner where the model adapts to the fixed apriori or fixed parameterized biases, meta learners dynamically choose the right biases [26].

The meta-learning tends to transfer knowledge learned from different environments to learn a new task with the least training as opposed to data-hungry supervised learning heavily biased due to i.i.d assumption. The evolution of transfer learning may help the machine to achieve system II cognition as of the human. The working principle of Meta-learning algorithm is presented below.

Step 1: Make a set of prior known tasks: $t_j \in T$

Step 2: Make a set of configurations resulted from the learning task $t_j$ such as hyperparameter settings, network architecture, and so on : $\theta_i \in \Theta$

Step 3: Prior evaluation measures (accuracy, FAR, training time, cross-validation ) of each configuration $\theta_i$ to task $t_j$: $P_{i,j}(\theta_i, t_j)$

Step 4: Assign a set of all prior scalar evaluations $P_{i,j}(\theta_i, t_j)$ of configuration $\theta_i$ on task $t_j$ to P: P= $\{P_{i,j}(\theta_i, t_j)\}$

Step 5: evaluate the performance $P_{i,new}$ on new task $t_{new}$ and assign it to $P_{new}$: $P_{new} = \{P_{i,new}\}$

Step 6: Now the meta-learner L is trained on $P'$ to predict recommended configurations $\Theta^*_{new}$ for new task $t_{new}$, where $P' = P \cup P_{new}$

Step 7: L is the learning algorithm derived from meta-learning to learn a new task

## 5.2. FEDERATED LEARNING

Federated learning (FL) is a machine learning framework where multiple nodes collaboratively train a model with local data under the orchestration of the centralized service provider/server [27]. Each node does not transfer or share locally stored data instead transfers the focused updates for immediate aggregation. In this way, a learning model can harness the privacy, security, regulatory and economic benefits [28]. In FL, we define N set of CAVs ready to collaborate $\{V_1, ...., V_N\}$ from different vendors with corresponding decentralized and isolated data $\{D_1, ...., D_N\}$. The conventional ML/DL model pulls up all the data $D = D_1 \cup ... \cup D_N$ and train the learning model L with D. In contrast, FL does not pulls up data D instead it share some model inference/parameters to train the learning model L.

The future CAV industry is envisioned with numerous CAVs from multiple vendors. With the lack of fully developed protocol standards, cross-vendors trust issues discourage data sharing among competing vendors. The various FL application in the CAV domain can be found in [29]–[32]. Article [29] describes how a piece of falsified information from a single CAV could disrupt the training of the global model. [30] proposed the dynamic federated proximal (DFP) based FL framework for designing the autonomous controller of the CAV. DFP is said to account for the mobility of CAVs, wireless fading channels, non-iid and unbalanced data. While solving the privacy leakage problem, FL has some inherent vulnerabilities such as model inversion, membership inference, and so on. [31] proposed Byzantine- fault-tolerant (BFT) decentralized FL with privacy preservation in CAV environment. Blockchain-based FL for CAV operations is proposed in [32]. Non-iid data distributions among multiple nodes, unbalanced datasets, communication latency are some of the challenges being solved in FL [27], [28], [33].

# 6. END TO END DEEP CNN-LSTM ARCHITECTURE FOR CAV CYBERATTACK DETECTION

In this chapter, we propose the novel deep CNN-LSTM architecture for attack detection and classification in the CAV environment as per Figure 5. Generally, CNN can learn the high dimensional spatial information of the feature space and may fail to capture distant temporal correlation. LSTM on other hand can capture the temporal correlation by learning the sequence. Thus, the stacked model can learn the Spatio-temporal features of the learning problem. The CNN-LSTM model is expected to perform better by learning hierarchical feature representations as well as learning long-term temporal dependencies of the huge data.

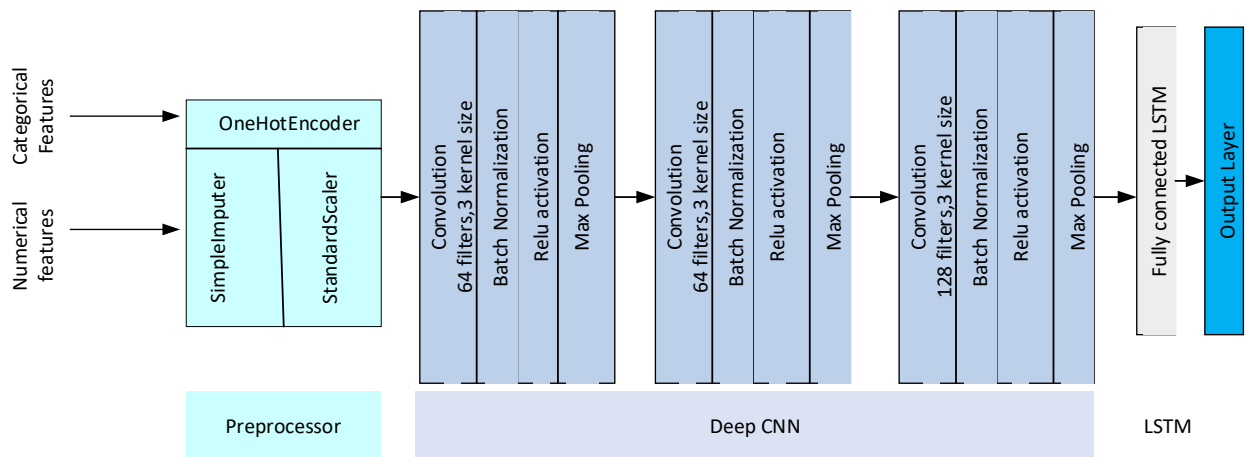

**FIGURE 5. Proposed end-to-end deep CNN-LSTM architecture.**

The proposed end-to-end Deep CNN-LSTM architecture pipeline has a preprocessor, a deep CNN layer, a fully connected LSTM, and an output layer.

a) Preprocessor: The preprocessor transforms the features into a machine learning compatible format. Most of the dataset contains the mixed type of features such as numerical (integer, float) and categorical (nominal, ordinal) datatypes. Deep learning algorithms perform the computation only in integer or float features. Therefore, all the categorical features should be transformed into numerical forms. OneHotEncoder transforms the nominal features into binary formats. However, the high cardinality features would be encoded with elevated dimensionality. For numerical features in the dataset, the SimpleImputer transformer deals with the missing values. Moreover, the standard scaler standardizes the features by implementing zero mean and unit variance. Finally, the preprocessor outputs the features ready to fetch to the Deep CNN architecture.

b) Deep CNN architecture: This layer is generally implemented to extract the spatial information using its kernel and present the high-level features. Being able to capture the local patterns, 1D CNN is the popular algorithm for time series classification/regression successfully tested in the fields of natural language processing, audio industry, and anomaly detection [34]. The *1D convolution layer* creates 64 convolution kernels of size three that convolve with the inputs over a single spatial or temporal dimension. The filters determine the dimensionality of output space while the kernel determines the length of the 1D convolution window. The *Batch Normalization Layer* normalizes the filter's output using the mean and standard deviations of the current batch of inputs. The activation function used is *ReLu* for solving the exploding and vanishing gradient of its other compatriots such as sigmoid, tanh [35]. *MaxPolling1D* downsamples the normalized filters' output by taking the maximum value over the spatial window of pool size four that extracts the high-level features. This marks the completion of a single block of deep CNN architecture, and we have added a similar block twice to extract more high-level features.

c) Fully connected LSTM: The 1D CNN generally extracts the local temporal information and is hard to capture all the long-term sequential correlations. That is where fully connected *LSTM* comes in handy to capture long-term sequential relations. The details of the LSTM model are explained in our previous journal work [36]. The LSTM layer has 64 LSTM units.

d) Output layer: The output layer has three nodes-belonging to three different classes- to evaluate the probability of sample belonging to each class. The probability sums to one with the highest probability indicating the predicted class taken care of by *softmax* [37] activation function. For one-hot encoded output classes, *categorical cross-entropy* [38] is used.

## 6.1. PERFORMANCE ANALYSIS
### 6.1.1. Dataset

The CAV-KDD dataset is adapted from KDD99 [39] dataset which is a well-known benchmark for intrusion detection. KDD99 dataset includes normal connections data and simulated attack data in a military network environment. Authors of [5] adapted by using 10% of KDD99 train data and 10% of KDD99 test data based to form the CAV-KDD train and test dataset. The train data and test data are mutually exclusive meaning the model has never experience the test data during model fitting. There are three kinds of data, normal data refers to the normal packets, Neptune and Smurf refer to the simulated DoS attacks. The reason for choosing only these three types is because deep learning models are data-hungry and need huge sample data to capture the distribution of the dataset. Further, we preprocess and refine the dataset that is going to be implementable in a deep learning environment. Table II. represents the distribution of the dataset while training and testing the CNN-LSTM. 30 % of training data is held for the validation of the model for the hyperparameter tuning. Table I indicates the class imbalance in the CAV train—20% belonging to the Normal category, 22% belonging to Neptune attack, and 58% belonging to Smurf attack – that inherently induces data biases in the learning model. Figure 6. presents the variance captured by the singular values over the 20 samples which in turn interpreted as the information captured by the prominent features. The four singular values i.e. four prime features can contribute up to 92.90% of data variance. Singular value decomposition (SVD) is the popular dimensionality reduction technique that projects the m-dimensional data (m-columns/features) into a subspace with m or fewer dimensions without losing the essence of original data [40].

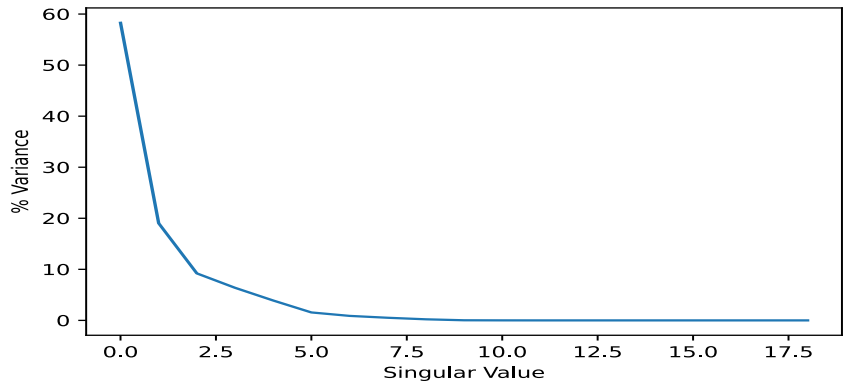

**FIGURE 6. Variance captured by the Singular values.**

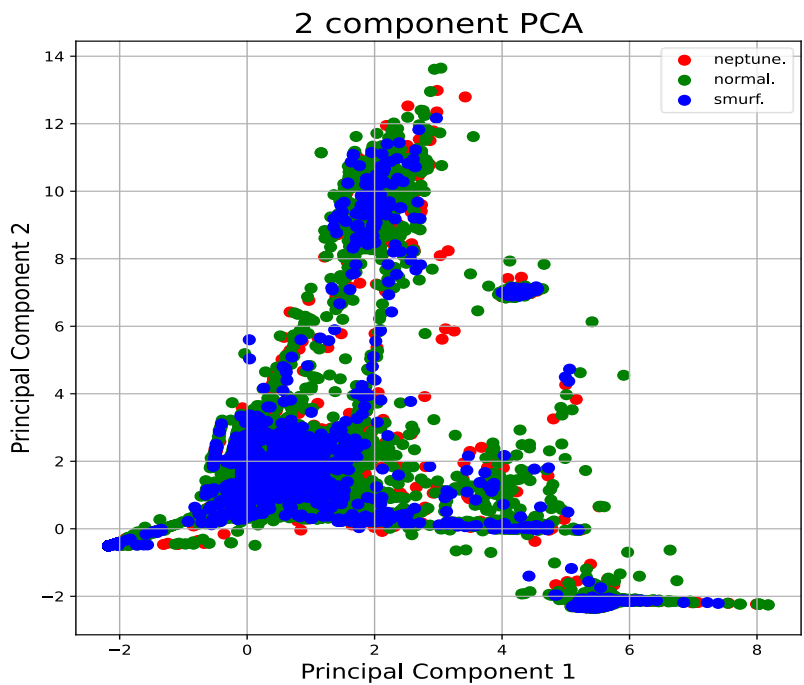

**FIGURE 7. Principal Component Analysis.**

**TABLE II.    DATA DISTRIBUTION**

| Data/Attack Label | CAV Train | CAV Test |
|---|---|---|
| Normal | 97,262 | 60,590 |
| Neptune | 107,201 | 58,001 |
| Smurf | 280,790 | 164,091 |

Principal component analysis (PCA) is the dimensionality reduction technique that uses the SVD to project data from hyperspace to lower dimensional space and extracts the dominant patterns in the matrix [41]. Figure 7. presents the PCA with two principal components over the 90,000 data samples showing tightly overlapped subspaces. In this notion, it's hard for any linear classifier to draw the non-linear boundaries between different classes. Therefore, a non-linear classifier such as deep learning could be handy. Deep learning is handy when one expects minimal or no feature selections since it can make good decisions with hyperdimensional feature space.

### 6.1.2. Evaluation metrics

The end-to-end deep CNN-LSTM architecture uses accuracy, precision, recall, F1-score to assess the performance of the classifier model. In this work, to quantify the performance of the proposed detection method, some performance metrics have been considered, such as accuracy, precision, recall, and F-1 score (defined below) from the confusion matrix. The confusion matrix generally reflects how efficiently a particular machine/algorithm classifies the actual data. It is the most ubiquitous matrix for the performance evaluation of the classifier, which is shown in table III., where the meaning of TP, FP, FN, and TN are described below.

**TABLE III. CONFUSION MATRIX**

| Actual class↓\Predicted class→ | Anomaly | Normal |
|---|---|---|
| Anomaly | TP | FN |
| Normal | FP | TN |

- True positive(TP): correctly classified intrusion,
- False-positive(FP): non-intrusive behavior wrongly classified as an intrusion,
- False-negative(FN): intrusive behavior wrongly classified as non-intrusive,
- True negative(TN): correctly classified non-intrusive behavior.

**Accuracy**: it estimates the correctly classified data out of all datasets. The higher the accuracy, the better the ML model. ($Accuracy \in= [0,1]$)

$$Accuracy = \frac{TP + TN}{TP + TN + FP + FN} \quad (1)$$

**Precision:** it estimates the ratio of correctly classified attacks to the number of all identified attacks. Precision represents the repeatability and reproducibility of the model ($Precision \in= [0,1]$). The higher the precision, the better the ML model.

$$Precision = \frac{TP}{TP + FP} \quad (2)$$

**True positive rate/Recall:** It estimates the ratio of a correctly classified anomaly to all anomaly data. A higher value is desired to be a better ML model and is given by: ($Recall \in= [0,1]$)

$$Recall = \frac{TP}{TP + FN} \qquad (3)$$

**F1-Score/Measure:** It is the harmonic mean of precision and recall. A higher value of F1-score represents the good ML model ($F1 - score \in= [0,1]$) and given by

$$F1 - score = 2 * \frac{Precision * Recall}{Precison + Recall} \qquad (4)$$

**AUC:** Area under the curve (AUC) tells the model's degree of separability. Higher the AUC better the model's separability.

## 6.2. RESULTS AND DISCUSSIONS

The experiments have been carried out in Intel® Core ™ i7 2.6 GHz CPU with 16 GB RAM computer. The Anaconda Navigator 2.0.4 hosts the JupyterLab 3.0.14 where algorithms are written in Python 3.8.8 of notebook 6.3.0. Our end-to-end deep CNN-LSTM model architecture is clearly explained including the number and size of convolution filters, kernel size, numbers and size of pooling layers, batch normalization layers, fully connected LSTM layer, and output layer as in section 6. For the comparison purpose, along with the CNN-LSTM, we created Deep Neural Network (DNN), Convolution Neural Network (CNN), Long-Short Term Memory (LSTM). The performance metrics of all the models are evaluated under similar constraints such as same train and test data, hyperparameters, batch size, and so on. Apart from that, all our deep learning models run for 10 epochs taking a batch of size 500 while training and all the models are tested with a batch size of 20. 30 % of the train data has been held out for validation so that one can tune the hyperparameters. The trained model has never experienced the features from the test data during the training so that the models don't over-parameterize and memorize.

Figure 8. presents the progression of training and validation accuracies as well as training and validation losses along the epochs. The training was so smooth that within 680 steps of the first epoch our proposed model achieves successive training and validation accuracies of 91.80% and 99.98% with successive losses of 0.3231 and 0.0052. The average training time per epoch was 95.1 seconds. The best model from the training has achieved 99.99% accuracy on validation data. This model can be trained up to just two epochs to get more than 99% training and validation accuracies.

Table IV. compares the precision, recall, f1-score, AUC, and testing accuracy of different deep learning algorithms against the proposed CNN-LSTM. The proposed CNN-LSTM model has achieved the highest precision, recall, f1-score, AUC, and testing accuracy i.e. more than 99% in each metrics among the class of other implemented algorithms. 1D CNN algorithm has achieved almost similar AUC and accuracy as the proposed CNN-LSTM algorithm. For similar setups with two hidden layers and 64 hidden nodes, LSTM is found to be the most inferior in terms of all other performance metrics except the AUC. All the performance metrics for DNN, CNN, CNN-LSTM were found to perform excellently with more than 99% evaluation metrics. This justifies the superior performance of our proposed deep CNN-LSTM algorithm. Similarly, Table V. presents the classwise performance of the proposed model where the proposed model exhibits almost 100% precision score for the samples from all three classes. The resulted recall and f1-score is almost 100% for Smurf and Neptune with 99% for samples from a normal class.

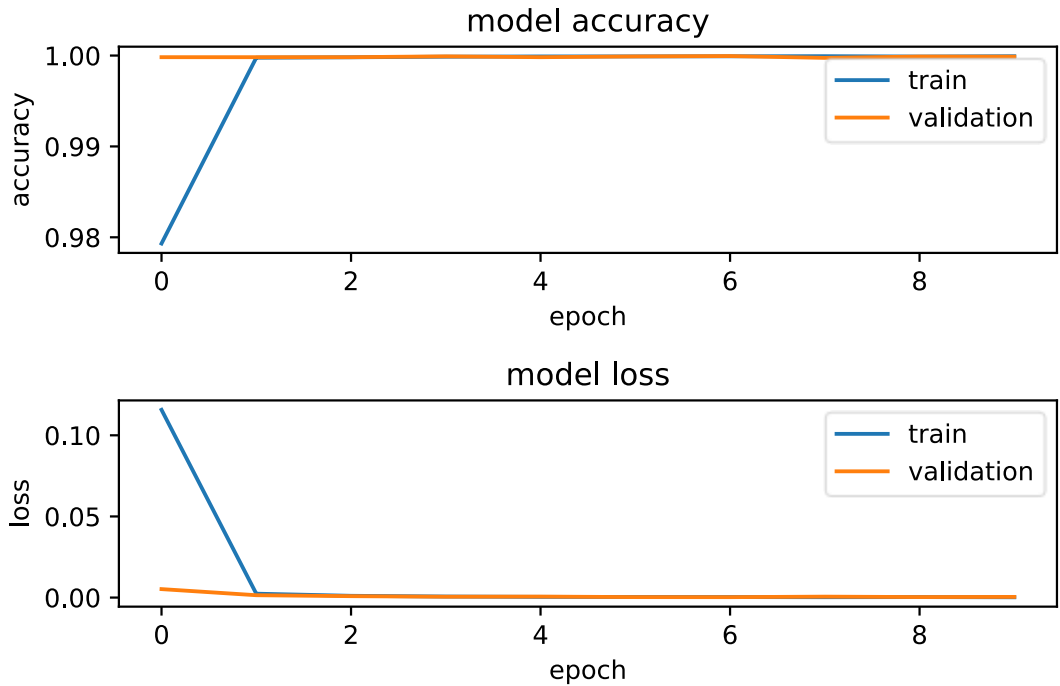

**FIGURE 8.** Training and Validation progression of deep CNN-LSTM.

**TABLE IV.** PERFORMANCE METRICS

| Algorithms | Precision | Recall | F1-score | AUC | Accuracy |
|---|---|---|---|---|---|
| DNN | 99.70% | 99.50% | 99.60% | 99.82% | 99.64% |
| CNN | 99.84% | 99.62% | 99.73% | 99.99% | 99.75% |
| LSTM | 92.41% | 96.29% | 93.85% | 99.95% | 93.89% |
| CNN-LSTM | 99.85% | 99.73% | 99.74% | 99.99% | 99.75% |

**TABLE V.** CLASSWISE PERFORMANCE EVALUATION OF THE PROPOSED DCNN-LSTM MODEL

| Label | Precision | Recall | F1-score | Support |
|---|---|---|---|---|
| Smurf | 1.00 | 1.00 | 1.00 | 164091 |
| Normal | 1.00 | 0.99 | 0.99 | 60590 |
| Neptune | 1.00 | 1.00 | 1.00 | 58001 |

Figure 9. presents the multiclass confusion matrix where each block has the number of samples with a percentage belonging to that block. The last row indicates the actual samples belonging to that classes while the last column represents the predicted samples using the proposed algorithm. The numbers and percentages in reds are misclassified samples. The highest misclassification rate of the proposed is 0.23% for the Smurf attack which is 640 out of 164,091 samples. This implicit misclassification i.e. bias came from the data distribution because smurf got almost 58.04% of total samples for the testing model got similar bias because of similar data distribution in the training. This bias in classifying Smurf resulted in false alarm of normal class i.e. there is a smurf attack but the model will predict it as a normal event. However, this error is less than 0.23% which is very small. But for high sensitivity CAV attack detection,

upsampling and downsampling can help to get equal data distribution while training the model. Overall, the proposed model has outstanding performance metrics almost close to 100%.

Confusion matrix

|  | Neptune | Normal | Smurf | sum_lin |
|---|---|---|---|---|
| Neptune | 57984 20.51% | 23 0.01% | 0 0.0% | 58007 99.96% 0.04% |
| Normal | 17 0.01% | 59927 21.20% | 24 0.01% | 59968 99.93% 0.07% |
| Smurf | 0 0.0% | 640 0.23% | 164067 58.04% | 164707 99.61% 0.39% |
| sum_col | 58001 99.97% 0.03% | 60590 98.91% 1.09% | 164091 99.99% 0.01% | 282682 99.75% 0.25% |

Predicted (y-axis), Actual (x-axis)

**FIGURE 9.** Confusion Matrix of deep CNN-LSTM

## 7. Conclusion

Along with the luxury of automation and connectivity, CAV inherits most of the cyber-physical vulnerabilities of incumbent technologies that primarily include evolving network architectures, wireless communications, and AI-assisted automation. This chapter sheds light on cyber-physical vulnerabilities and risks that originated in IT, OT, and the physical domains of the CAV ecosystem, eclectic threat landscapes, and threat intelligence. To deal with the security threats embedded in high-speed, high dimensional, multimodal data and assets from eccentric stakeholders of the CAV ecosystem, this chapter presents and analyzes some of the state of art deep learning-based threat intelligence for attack detection. Since deep learning itself has been evolving to attain superior cognition and intelligence, it would have a direct impact on threat intelligence as well. The collaborative learning platform of deep learning, federated learning, is capable of sharing threat intelligence without the need to share the data between divergent stakeholders of the CAV ecosystem. Also, deep learning for CAV has still to work on Meta-learning for robust and swift generalization under the dynamic environment and out of distribution data context. The frontiers in deep learning along with the challenges have been included in the chapter. We have proposed, trained, and tested the deep CNN-LSTM model for CAV threat intelligence; assessed and compared the performance of the proposed model against other deep learning algorithms such as DNN, CNN, LSTM. Our results indicate the superiority of the proposed model although DNN and 1d-CNN also achieved more than 99% of accuracy, precision, recall, f1-score, and AUC on the CAV-KDD dataset. The outstanding performance of deep CNN-LSTM comes with the increased model complexity and cumbersome hyperparameters tuning. Still, there are open challenges on deep learning adoption in the CAV cybersecurity paradigm due to costlier implementations and training, lack of properly developed protocols and policies, poorly defined privileges between stakeholders, adversarial threats to the deep learning model, and poor generalizability of the model under out of data distributions.

# REFERENCES

[1] "Motor Vehicle Deaths in 2020 Estimated to be Highest in 13 Years, Despite Dramatic Drops in Miles Driven - National Safety Council." https://www.nsc.org/newsroom/motor-vehicle-deaths-2020-estimated-to-be-highest (accessed Jul. 21, 2021).

[2] "Preliminary Monthly Estimates," *Injury Facts*. https://injuryfacts.nsc.org/motor-vehicle/overview/preliminary-monthly-estimates/ (accessed Aug. 25, 2021).

[3] F. Falcini and G. Lami, "Deep Learning in Automotive: Challenges and Opportunities," in *Software Process Improvement and Capability Determination*, vol. 770, A. Mas, A. Mesquida, R. V. O'Connor, T. Rout, and A. Dorling, Eds. Cham: Springer International Publishing, 2017, pp. 279–288. doi: 10.1007/978-3-319-67383-7_21.

[4] "Security Challenges for Connected and Autonomous Vehicles," *BAE Systems | Cyber Security & Intelligence*. https://www.baesystems.com/en/cybersecurity/feature/security-challenges-for-connected-and-autonomous-vehicles (accessed Sep. 12, 2021).

[5] Q. He, X. Meng, R. Qu, and R. Xi, "Machine Learning-Based Detection for Cyber Security Attacks on Connected and Autonomous Vehicles," *Mathematics*, vol. 8, no. 8, p. 1311, Aug. 2020, doi: 10.3390/math8081311.

[6] "Autonomous cars generate more than 300 TB of data per year," *Tuxera*, Jul. 02, 2021. https://www.tuxera.com/blog/autonomous-cars-300-tb-of-data-per-year/ (accessed Jul. 21, 2021).

[7] "J3016C: Taxonomy and Definitions for Terms Related to Driving Automation Systems for On-Road Motor Vehicles - SAE International." https://www.sae.org/standards/content/j3016_202104/ (accessed Aug. 24, 2021).

[8] J. Ashraf, A. D. Bakhshi, N. Moustafa, H. Khurshid, A. Javed, and A. Beheshti, "Novel Deep Learning-Enabled LSTM Autoencoder Architecture for Discovering Anomalous Events From Intelligent Transportation Systems," *IEEE Trans. Intell. Transp. Syst.*, vol. 22, no. 7, pp. 4507–4518, Jul. 2021, doi: 10.1109/TITS.2020.3017882.

[9] "Automated Vehicles for Safety | NHTSA." https://www.nhtsa.gov/technology-innovation/automated-vehicles-safety (accessed Aug. 25, 2021).

[10] S. Parkinson, P. Ward, K. Wilson, and J. Miller, "Cyber Threats Facing Autonomous and Connected Vehicles: Future Challenges," *IEEE Trans. Intell. Transp. Syst.*, vol. 18, no. 11, pp. 2898–2915, Nov. 2017, doi: 10.1109/TITS.2017.2665968.

[11] A. M. Wyglinski, X. Huang, T. Padir, L. Lai, T. R. Eisenbarth, and K. Venkatasubramanian, "Security of Autonomous Systems Employing Embedded Computing and Sensors," *IEEE Micro*, vol. 33, no. 1, pp. 80–86, Jan. 2013, doi: 10.1109/MM.2013.18.

[12] N. Statt, "Self-driving car engineer Anthony Levandowski pleads guilty to stealing Google trade secrets," *The Verge*, Mar. 19, 2020. https://www.theverge.com/2020/3/19/21187651/anthony-levandowski-pleads-guilty-google-waymo-uber-trade-secret-theft-lawsuit (accessed Sep. 12, 2021).

[13] Z. El-Rewini, K. Sadatsharan, D. F. Selvaraj, S. J. Plathottam, and P. Ranganathan, "Cybersecurity challenges in vehicular communications," *Veh. Commun.*, vol. 23, p. 100214, Jun. 2020, doi: 10.1016/j.vehcom.2019.100214.

[14] J. Shen, J. Y. Won, Z. Chen, and Q. A. Chen, "Demo: Attacking Multi-Sensor Fusion based Localization in High-Level Autonomous Driving," in *2021 IEEE Security and Privacy Workshops (SPW)*, May 2021, pp. 242–242. doi: 10.1109/SPW53761.2021.00039.

[15] "Cybersecurity for Connected and Autonomous Vehicles," p. 36, 2019.

[16] "AAAI 2020 Conference | Thirty-Fourth AAAI Conference on Artificial Intelligence." https://aaai.org/Conferences/AAAI-20/ (accessed Sep. 12, 2021).

[17] Y. Bengio, Y. Lecun, and G. Hinton, "Deep learning for AI," *Commun. ACM*, vol. 64, no. 7, pp. 58–65, Jun. 2021, doi: 10.1145/3448250.

[18] ICML IJCAI ECAI 2018 Conference Videos, *AAAI 20 / AAAI 2020 Keynotes Turing Award Winners Event / Geoff Hinton, Yann Le Cunn, Yoshua Bengio*, (Feb. 10, 2020). Accessed: Sep. 12, 2021. [Online Video]. Available: https://www.youtube.com/watch?v=UX8OubxsY8w

[19] M. Basnet, S. Poudyal, M. H. Ali, and D. Dasgupta, "Ransomware Detection Using Deep Learning in the SCADA System of Electric Vehicle Charging Station," *ArXiv210407409 Cs Eess*, Apr. 2021, Accessed: Jun. 04, 2021. [Online]. Available: http://arxiv.org/abs/2104.07409

[20] A. Kavousi-Fard, M. Dabbaghjamanesh, T. Jin, W. Su, and M. Roustaei, "An Evolutionary Deep Learning-Based Anomaly Detection Model for Securing Vehicles," *IEEE Trans. Intell. Transp. Syst.*, vol. 22, no. 7, pp. 4478–4486, Jul. 2021, doi: 10.1109/TITS.2020.3015143.

[21] E. Seo, H. M. Song, and H. K. Kim, "GIDS: GAN based Intrusion Detection System for In-Vehicle Network," *2018 16th Annu. Conf. Priv. Secur. Trust PST*, pp. 1–6, Aug. 2018, doi: 10.1109/PST.2018.8514157.

[22] H. M. Song, J. Woo, and H. K. Kim, "In-vehicle network intrusion detection using deep convolutional neural network," *Veh. Commun.*, vol. 21, p. 100198, Jan. 2020, doi: 10.1016/j.vehcom.2019.100198.

[23] J. J. Q. Yu, "Sybil Attack Identification for Crowdsourced Navigation: A Self-Supervised Deep Learning Approach," *IEEE Trans. Intell. Transp. Syst.*, vol. 22, no. 7, pp. 4622–4634, Jul. 2021, doi: 10.1109/TITS.2020.3036085.

[24] D. Kahneman, *Thinking, Fast and Slow*, 1st edition. New York: Farrar, Straus and Giroux, 2013.

[25] J. Vanschoren, "Meta-Learning: A Survey," *ArXiv181003548 Cs Stat*, Oct. 2018, Accessed: Aug. 30, 2021. [Online]. Available: http://arxiv.org/abs/1810.03548

[26] R. Vilalta and Y. Drissi, "A Perspective View and Survey of Meta-Learning," p. 20.

[27] P. Kairouz *et al.*, "Advances and Open Problems in Federated Learning," *ArXiv191204977 Cs Stat*, Mar. 2021, Accessed: Aug. 31, 2021. [Online]. Available: http://arxiv.org/abs/1912.04977

[28] Y. Zhao, M. Li, L. Lai, N. Suda, D. Civin, and V. Chandra, "Federated Learning with Non-IID Data," *ArXiv180600582 Cs Stat*, Jun. 2018, Accessed: Aug. 31, 2021. [Online]. Available: http://arxiv.org/abs/1806.00582

[29] R. A. Mallah, G. Badu-Marfo, and B. Farooq, "Cybersecurity Threats in Connected and Automated Vehicles based Federated Learning Systems," *ArXiv210213256 Cs*, Jun. 2021, Accessed: Aug. 31, 2021. [Online]. Available: http://arxiv.org/abs/2102.13256

[30] T. Zeng, O. Semiari, M. Chen, W. Saad, and M. Bennis, "Federated Learning on the Road: Autonomous Controller Design for Connected and Autonomous Vehicles," *ArXiv210203401 Cs Eess*, Feb. 2021, Accessed: Aug. 31, 2021. [Online]. Available: http://arxiv.org/abs/2102.03401

[31] J.-H. Chen, M.-R. Chen, G.-Q. Zeng, and J. Weng, "BDFL: A Byzantine-Fault-Tolerance Decentralized Federated Learning Method for Autonomous Vehicles," *IEEE Trans. Veh. Technol.*, pp. 1–1, 2021, doi: 10.1109/TVT.2021.3102121.

[32] Y. Fu, F. R. Yu, C. Li, T. H. Luan, and Y. Zhang, "Vehicular Blockchain-Based Collective Learning for Connected and Autonomous Vehicles," *IEEE Wirel. Commun.*, vol. 27, no. 2, pp. 197–203, Apr. 2020, doi: 10.1109/MNET.001.1900310.

[33] M. Mohri, G. Sivek, and A. T. Suresh, "Agnostic Federated Learning," in *International Conference on Machine Learning*, May 2019, pp. 4615–4625. Accessed: Aug. 31, 2021. [Online]. Available: https://proceedings.mlr.press/v97/mohri19a.html

[34] W. Tang, G. Long, L. Liu, T. Zhou, J. Jiang, and M. Blumenstein, "Rethinking 1D-CNN for Time Series Classification: A Stronger Baseline," *ArXiv200210061 Cs Stat*, Feb. 2021, Accessed: Sep. 07, 2021. [Online]. Available: http://arxiv.org/abs/2002.10061

[35] M. Basnet and M. H. Ali, "Deep Learning-based Intrusion Detection System for Electric Vehicle Charging Station," in *2020 2nd International Conference on Smart Power Internet Energy Systems (SPIES)*, Sep. 2020, pp. 408–413. doi: 10.1109/SPIES48661.2020.9243152.

[36] "Exploring cybersecurity issues in 5G enabled electric vehicle charging station with deep learning - Basnet - - IET Generation, Transmission & Distribution - Wiley Online Library." https://ietresearch.onlinelibrary.wiley.com/doi/10.1049/gtd2.12275 (accessed Sep. 08, 2021).

[37] A. Martins and R. Astudillo, "From Softmax to Sparsemax: A Sparse Model of Attention and Multi-Label Classification," in *International Conference on Machine Learning*, Jun. 2016, pp. 1614–1623. Accessed: Sep. 08, 2021. [Online]. Available: https://proceedings.mlr.press/v48/martins16.html

[38] Z. Zhang and M. Sabuncu, "Generalized Cross Entropy Loss for Training Deep Neural Networks with Noisy Labels," in *Advances in Neural Information Processing Systems*, 2018, vol. 31. Accessed: Sep. 08, 2021. [Online]. Available: https://proceedings.neurips.cc/paper/2018/hash/f2925f97bc13ad2852a7a551802feea0-Abstract.html

[39] "KDD Cup 1999 Data." http://kdd.ics.uci.edu/databases/kddcup99/kddcup99.html (accessed Sep. 08, 2021).

[40] M. E. Wall, A. Rechtsteiner, and L. M. Rocha, "Singular Value Decomposition and Principal Component Analysis," in *A Practical Approach to Microarray Data Analysis*, D. P. Berrar, W. Dubitzky, and M. Granzow, Eds. Boston, MA: Springer US, 2003, pp. 91–109. doi: 10.1007/0-306-47815-3_5.

[41] S. Wold, K. Esbensen, and P. Geladi, "Principal component analysis," *Chemom. Intell. Lab. Syst.*, vol. 2, no. 1, pp. 37–52, Aug. 1987, doi: 10.1016/0169-7439(87)80084-9.